\documentclass[aps,prl,twocolumn,showpacs,amsmath,amssymb,amsfonts,floatfix]{revtex4}

\usepackage[pdftex]{graphicx}
\usepackage{color}
\usepackage{textcomp}

\renewcommand{\section}[1]{\textit{#1} --}

\begin{document}

\title{A Concept of Linear Thermal Circulator Based on Coriolis forces}
\author{Huanan Li$^{1}$ and Tsampikos Kottos$^{1,2}$}
\affiliation{\mbox{$^1$Department of Physics, Wesleyan University, Middletown, Connecticut 06459, USA}\\
\mbox{$^2$Max Planck Institute for Dynamics and Self-organization (MPIDS),  37077 G\"{o}ttingen, Germany}}
\begin{abstract}
We show that the presence of a Coriolis force in a rotating linear lattice imposes a non-reciprocal propagation of the phononic 
heat carriers. Using this effect we propose the concept of Coriolis linear thermal circulator which can control the circulation of a 
heat current. A simple model of three coupled harmonic masses on a rotating platform allow us to demonstrate giant circulating 
rectification effects for moderate values of the angular velocities of the platform.
\end{abstract}

\maketitle
\section{Introduction} Directional transport and the creation of non-reciprocal devices that control the flow of energy and/or 
mass at predefined directions, have been posing always fascinating challenges for both theoretical physicists and engineers 
\cite{K00,MEW13,C45}. On the theoretical side, the main difficulty is to find ways to bypass time-reversal symmetry that many 
linear systems exhibit and which ensures reciprocal transmission. At the same time the current technological needs for high-
performance enduring on-cheep integrated devices dictates certain limitations on the realization of such directional valves 
which constitute the basic building blocks for a variety of devices ranging from rectifiers and circulators, to pumps, switches 
and transistors. 

Despite the various challenges, in recent years many non-reciprocal structures for general wave flows have been theoretically 
proposed and subsequently engineered in contexts as diverse as photonics, acoustics and thermal transport. For example 
in the framework of photonics, optical diodes are mainly based on magneto-optical phenomena like the Faraday effect caused
by non-reciprocal circular birefringence. An alternative pathway for directional photonic transport is the use of non-linear 
elements which in the presence of asymmetric scattering potentials \cite{GA99}, active elements \cite{BFBRCEK13,CJHYWJLWX14,
POLMGLFNBY14} etc. can induce strong asymmetric transport. The use of non-linearities for the creation of one-way valves 
was proven successful also in acoustics, see for example Ref. \cite{BTD11}. However nonlinear mechanisms often introduce 
inherent signal distortions (higher harmonics generation) and also they impose limitations on the operational amplitude of 
the device - an undesirable feature from the engineering perspective. Finally, unidirectional sound propagation with linear 
components has been also reported in Refs. \cite{XL11}, but the non-reciprocal response is either weak or the real estate 
needed to observe considerable non-reciprocal effects has to be large. A recent breakthrough in this direction was reported 
in Ref. \cite{FSSHA14} where it was demonstrated that by employing an acoustic analogue of Zeeman effect one can achieve 
giant linear non-reciprocity in compact structures.

\begin{figure}
\includegraphics[width=0.75\linewidth, angle=0]{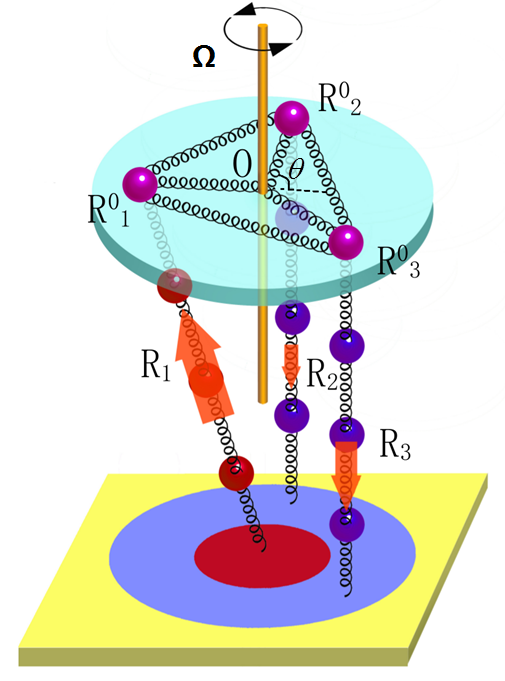}
\caption{(Color online) A schematic of the minimum linear rectifier: Three equal masses coupled together with equal harmonic
springs. The masses are also attached to a post with similar springs and they move on a platform at the $X-Y$ plane which rotates
with a counterclockwise (CCW) angular velocity $\Omega$. The mass $R_1$ is coupled to a bath with temperature $T_{R_1}=T_H$ 
while the other two masses $R_2,R_3$ are coupled to reservoirs with equal temperatures $T_{R_2}=T_{R_3}=T_L<T_H$. The heat current 
flowing towards the reservoir $R_3$ is rectified with respect to the current flowing into reservoir $R_2$ despite the geometric symmetry
of our structure.
\label{fig1}}
\end{figure}

The endeavor for unidirectional devices carried over also to phonons as carriers of heat energy. Indeed Casati and collaborators 
have proposed a thermal rectification mechanism that relies on nonlinear lattice dynamics \cite{TPC02}. A series of subsequent 
works which have been based on this idea have resulted in a wealth of new designs with better rectification characteristics (see review 
\cite{LRWZHL12}). The theoretical efforts have been culminated with the work of Ref. \cite{COMZ06} where a first experimental realization 
of a nanoscale thermal diode has been demonstrated. Despite this success, from the engineer perspective and for the same reasons 
as the ones discussed above, it is desirable to have linear nanoscale devices that produce strong thermal rectification. This goal has 
been proven far more challenging in the framework of thermal transport than in photonics or acoustics.

In this Letter we propose the concept of the Coriolis linear thermal circulators in analogy to their microwave counterparts. The structure 
consists of a linear phononic lattice attached to three reservoirs. One of the reservoirs is kept at high temperatute while the other two 
are kept at the same low temperature. The set-up is placed on a rotating platform with externally tunable angular velocity $\Omega$. 
We demonstrate the above concept using a simplified model consisting of three harmonic masses (see Fig.~\ref{fig1}). We find that a 
current flowing towards a predefined low-temperature bath is rectified as high as $\approx 90\%$ for moderate values of $\Omega$. 
The physical mechanism behind this rectification effect is related with the directional bias which is imposed to the linear system by the 
Coriolis force when it is acting on the rotating masses. 

{\it General formalism for rotating lattices-} Let us consider $N$ particles of equal masses forming a lattice which rotates with a constant 
angular velocity $\overrightarrow{\Omega}$ around the $Z$-axis. Up to the harmonic approximation, the Hamiltonian for this 
lattice in the rotating frame is
\begin{equation}
H_{C}= \frac{1}{2}p_{C}^{T}p_{C}+\frac{1}{2}S^{T}K^{C}S-\left(R^{0}+S\right)^{T}A\, p_{C}
\label{hamilt}
\end{equation}
where the superscript $T$ stands for matrix transpose. The vector $R^{0}=(R_1^0,\cdots,R_N^0)^T$ describes the equilibrium positions 
of the lattice particles $R_1,\cdots,R_N$
in the rotating frame, while the vectors $S=(S_1^x,S_1^y,S_1^z,\cdots,S_N^x,S_N^y,S_N^z)^T$ and $p_{C}=(p_{C1}^x,p_{C1}^y,p_{C1}^z,
\cdots,p_{CN}^x,p_{CN}^y,p_{CN}^z)^T$ describe their mass-reduced displacements and associated conjugate canonical momenta. The 
dimensionality of these vectors is ${\cal N}=N\cdot D$ where $D$ is the dimensionality of the space. 
The force matrix $K^{C}$ is ${\cal N}\times {\cal N}$. The last term in Eq. (\ref{hamilt}) describes the Coriolis force in the rotating frame. 
The matrix $A$ is an ${\cal N}\times {\cal N}$ block-diagonal matrix defined as
\begin{equation}
\label{Amatrix}
A={\rm diag}\{ {\tilde A}_D\}; {\tilde A}_{D=3}=
\begin{bmatrix} {\tilde A}_{D=2} & 0\\
0                                                 &0
\end{bmatrix}; {\tilde A}_{D=2}=
\begin{bmatrix} 0 & \Omega\\
 -\Omega            &0
\end{bmatrix}
\end{equation}
where for the $D=2$ case we have assumed a motion on the $X-Y$ plane.

Furthermore we assume that the rotating lattice Eq. (\ref{hamilt}) is connected with three equivalent co-rotating heat baths which are described 
quasi-classically i.e. we promote the relative momenta $p_{\alpha}$ of the $\alpha-$bath particles and their displacements $u_{\alpha}$ 
with respect to the rotating frame to conjugate canonical pairs. Note that although this treatment is not formally correct on the quantum
mechanical level it can, nevertheless, justified on the classical level (high bath temperatures) \cite{Landau1980}. Namely the bath Hamiltonians 
are
\begin{align}
H_{\alpha}= & \frac{1}{2}p_{\alpha}^{T}p_{\alpha}+\frac{1}{2}u_{\alpha}^{T}K^{\alpha}u_{\alpha},\, \alpha=R_1,R_2,R_3
\label{bath}
\end{align}
where the (semi-infinite) force matrix $K^{\alpha}$ contains additional $-\left|\overrightarrow{\Omega}\right|^{2}\cdot I$ terms due to the 
centrifugal force ($I$ denotes the semi-infinite identity matrix). In this treatment, the statistical properties of the heat baths are not affected 
by the Coriolis's force. The sub-index $\alpha$ denotes the heat-bath attached to particle $\alpha$. We will always assume that there are 
only three baths attached to three different particles of the lattice $R_1,R_2,R_3$ with temperatures $T_{R_1}=T_H>T_{R_2}=T_{R_3}=T_L$. 
Finally we note that in all force matrices $K^C, K^{\alpha}$ appearing in Eqs. (\ref{hamilt}, \ref{bath}) a quadratic pinning potential is introduced, 
that guarantees the existence of equilibrium positions for the lattice particles. This potential originates from the interaction between the 
rotating system and the substrate. Formally it is introduced as additional diagonal terms $k_{0}\cdot I$ in the force matrices.

The total Hamiltonian of the bath-lattice system is
\begin{equation}
H_{tot}= H_{C}+\sum_{\alpha}H_{\alpha}+\sum_{\alpha}H_{\alpha C}
\label{total}
\end{equation}
with $H_{\alpha C}=u_{\alpha}^{T}V^{\alpha C}u_{C}$
being the coupling between the lattice particles and the heat baths.

{\it Method -} We use nonequilibrium Green's function method (NEGF) to calculate the steady thermal current~\cite{Wang2013}. 
Specifically, the steady-state current out of the heat bath $\alpha$ is 
\begin{equation}
I_{\alpha}
=  \int_{0}^{\infty}\frac{d\omega}{2\pi}\hbar\omega\sum_{\gamma=R_{1},R_{2},R_{3}}{\cal T}_{\gamma\alpha}\left[\omega\right]
\left(f_{\alpha}-f_{\gamma}\right)
\label{eq:Landauer-like}
\end{equation}
with $f_{\alpha}=\left\{ \exp\left(\hbar\omega/k_B T_{\alpha}\right)/-1\right\} ^{-1}$ being the Bose-Einstein distribution for the 
heat bath $\alpha$, and ${\cal T}_{\gamma\alpha}\left[\omega\right]=\mathrm{Tr}\left[G_{CC}^{r}\Gamma_{\gamma}G_{CC}^{a}
\Gamma_{\alpha}\right]$ is the transmission coefficient from bath $\alpha$ to bath $\gamma$. The ${\cal N}$-dimensional matrix 
$\Gamma_{\alpha}\equiv i\left[\Sigma_{\alpha}^{r}-\Sigma_{\alpha}^{a}\right]$ can be easily calculated from the relation between 
the retarded/advanced ($r/a$) self-energy $\Sigma_{\alpha}^{r/a}$ and the corresponding equilibrium Green's function 
$g_{\alpha}^{r}$ of the isolated heat bath $\alpha$, $\mathit{i.e.}$, $\Sigma_{\alpha}^{r/a}=\left(V^{\alpha C}\right)^{T}
g_{\alpha}^{r/a}V^{\alpha C}$. Then the whole problem collapses to the study of the Green's functions of the lattice $G_{CC}^{r/a}$. 

We study the contour-ordered Green's function of the lattice $G_{CC}\left(\tau,\tau'\right)$ using the equation of motion 
method~\cite{Zhang2009}. The associated Dyson equation reads, 
\begin{widetext}
\begin{equation}
G_{CC}\left(\tau,\tau'\right)= g_{C}\left(\tau,\tau'\right)-\left(K^{C}\right)^{-1}A^{2}R^{0}G_{C}^{T}+\int_{C}d\tau_{1}d\tau_{2}g_{C}
\left(\tau,\tau_{1}\right)\left[\left(A^{2}-2A\frac{\partial}{\partial\tau_{1}}\right)\delta\left(\tau_{1},\tau_{2}\right)+\Sigma
\left(\tau_{1},\tau_{2}\right)\right]G_{CC}\left(\tau_{2},\tau'\right) .
\label{Dyson}
\end{equation}
\end{widetext}
Above, the contour variables $\tau$ are defined on the Keldysh contour $C$~\cite{Schwinger1961}, the generalized $\delta$-function 
$\delta\left(\tau_{1},\tau_{2}\right)$ is the counterpart of the ordinary Dirac delta function on the same contour $C$, $\Sigma=
\sum_{\alpha}\Sigma_{\alpha}$ denotes the total self-energy due to the interaction with all the heat baths, $g_{C}$ is the equilibrium 
Green's function for the isolated lattice and $G_C$ is the one-point Green's function in the steady state. Using the Langreth theorem
~\cite{Haug2008} and then Fourier transforming the obtained real-time Green's functions, we get various useful relations such as 
\begin{eqnarray}
G_{CC}^{<}\left[\omega\right]= & G_{CC}^{r}\left[\omega\right]\Sigma^{<}\left[\omega\right]G_{CC}^{a}
\left[\omega\right]-\nonumber\\&
\left(K^{C}\right)^{-1}A^{2}R^{0}
G_{C}^{T}2\pi\delta\left(\omega\right),
\label{eq:keystep}
\end{eqnarray}
where $G_{CC}^{<}\left[\omega\right]$ is the lesser Green's function. Equation (\ref{eq:keystep}) is critical in deriving 
Eq.~\eqref{eq:Landauer-like}. Finally, the retarded Green's function $G_{CC}^r$ is obtained from Eq. (\ref{Dyson}) and reads
\begin{equation}
G^{r}\left[\omega\right]
= \left[\left(\omega+i0^{+}\right)^{2}-K^{C}-\Sigma^{r}\left[\omega\right]-A^{2}-2i\omega A\right]^{-1}
\end{equation}
The associated advanced Green function is evaluated as $G^{a}[\omega]=\left(G^{r}\left[\omega\right]\right)^{\dagger}$. These 
expressions allow us to calculate the transmission coefficient ${\cal T}_{\gamma\alpha}[\omega]$ used in Eq. (\ref{eq:Landauer-like}).

{\it A minimum model-} Next we proceed with a demonstration of the rectification phenomenon in the presence of a Coriolis force using 
a simplified version of the general model Eq. (\ref{total}). The system that we consider consists of three equal masses coupled together 
with harmonic coupling $k^C$. Each mass is coupled with the same harmonic coupling to a post which is placed at position $O$. 
An additional coupling $k_0$ with the substrate is assumed. The particles are moving on a counterclockwise (CCW) rotating plane 
with angular velocity $\Omega$. The equilibrium configuration of the system is defined by the vector $R^0=(R_1^0,R_2^0,R_3^0)^T$ 
which can be parametrized in terms of an equilibrium angle $\theta$ such that $2\theta=\angle R_2^0OR_3^0; \pi-\theta=\angle 
R_1^0OR_2^0=\angle R_1^0OR_3^0$. 

One of the particles is attached to a 1D Rubin bath \cite{Rubin1971} which is kept at high temperature $T_{R_1}=T_H$ while the 
other two particles are attached to two other independent 1D Rubin baths with the same low temperature $T_{R_2}=T_{R_3}=T_L$. 
An illustration of our minimal model is shown in Fig.~\ref{fig1}. The 1D Rubin baths $\alpha=R_{1},R_{2},R_{3}$ are made up of a 
semi-infinite spring chain with $K^{\alpha}_{nm}=\delta_{nm} (2k^{\alpha}-\Omega^{2}+k_{0})-k^{\alpha}\delta_{n\pm1,m}$. In 
order to guarantee the stability of the heat bath we need to make sure that the magnitude of the angular velocity $\Omega$ is 
less than $\sqrt{k_{0}}$. A local coordinate system for each bath is set up for convenience. Then the force matrix $K^{C}$ is
$K^{C}=  k^{C}\begin{bmatrix}D^{R_1} & \Theta_{1}^{T} & \Theta_{1}\\
\Theta_{1} & D^{R_{2}} & \Theta_{2}\\
\Theta_{1}^{T} & \Theta_{2}^{T} & D^{R_{3}}
\end{bmatrix};
\Theta_{n}=\begin{bmatrix}-\cos(n\theta) & -\sin(n\theta)\\
\sin(n\theta) & -\cos(n\theta)
\end{bmatrix}
$
where $D^{\alpha}_{nm}=\left(3+\frac{k^{\alpha}}{k^{C}}+\frac{k_{0}}{k^{C}}\right)\delta_{nm}$ is a $2D$ sub-matrix. 
Finally, the nonzero elements of the coupling matrices appearing in Eq. (\ref{total}) are respectively $\left(V^{R_1C}\right)_{1,1}=-k_{R_1}$,
$\left(V^{R_{2}C}\right)_{1,3}=-k_{R_{2}}$ and $\left(V^{R_{3}C}\right)_{1,5}=-k_{R_{3}}$. In our analysis below we will 
assume for simplicity that $k^{\alpha}=k^C=k$.

We start out analysis with the investigation of the normal modes of the closed system (no bath attached). Substitution of the displacement 
vector $S=\exp(i\omega t) \cdot {\tilde S}$ into the Hamilton's equation ${\dot p}_C=-{\partial H_c\over \partial S}; {\dot S}= 
{\partial H_c\over \partial p_C}$ allow us to obtain the following equations of motion
\begin{equation}
\label{eqmot}
\left[(-\omega^2 I_{\cal N} + K^C) + (A-2i \omega)\cdot A\right]= - A^2 R^0
\end{equation} 
The normal modes are found by solving the secular equation associated with the homogeneous part of Eq. (\ref{eqmot}). We get
\begin{eqnarray}
\label{spectrum}
\omega_{1}=\omega_{2}=\sqrt{5k+k_{0}}+\Omega;&\,\,
\omega_{3}=\omega_{4}=\sqrt{5k+k_{0}}-\Omega; \nonumber \\
\omega_{5}=\sqrt{2k+k_{0}}+\Omega;&\,
\omega_{6}=\sqrt{2k+k_{0}}-\Omega
\end{eqnarray}
This spectrum bears strong analogies with the Zeeman effect where the role of the external magnetic field is now
played by the axial vector of angular momentum ${\vec \Omega}$. In the absence of rotation $\Omega=0$, corresponding to
$A=0$ in Eq. (\ref{eqmot}), the ground state has a double degeneracy while the excited state has a four-fold degeneracy. 
Once the rotation is introduced the degeneracies are lifted completely for the ground state and partially for the excited state. 
The additional degeneracy appearing for the excited state is due to the reflection symmetry of our set-up with respect to the 
axis $OR_1$.

The spectrum Eq. (\ref{spectrum}) can be further understood by using degenerate perturbation theory: one can decompose the terms 
appearing on the l.h.s of Eq. (\ref{eqmot}) to an unperturbed term involving the force matrix $K^C$ and a perturbation term 
involving $-2i\omega_0 A\sim \Omega$ where $\omega_0$ is the normal mode associated with the unperturbed system $\Omega=0$
(the $A^2$ term is irrelevant in the argumentation as it is proportional to the identity matrix). It turns out that in the normal 
mode basis, associated with the unperturbed system, the matrix $A$ maintains its block-diagonal form (see Eq. (\ref{Amatrix})), 
thus mixing only pairs of degenerate modes. It is then straightforward to see that the correction terms are proportional to $\pm \Omega$.

\begin{figure}
\includegraphics[width=1\linewidth,angle=0]{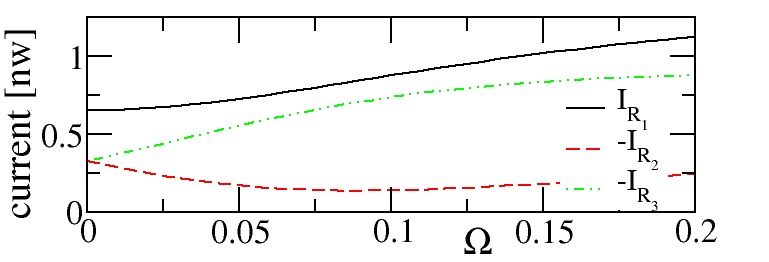}
\caption{(Color online) Plots of steady-state currents $I_{R_{1}}$,  $-I_{R_{2}}$ and $-I_{R_{3}}$ versus the angular velocity 
$\Omega \left[{\rm \frac{1}{\AA}\sqrt{\frac{eV}{u}}}\right]$ when  the configuration angle is $\theta=60^\circ$. Other parameters 
are $T_{R_1}=T_{H}=310\mathrm{K},\: T_{R_2}=T_{R_3}=T_{L}=290\mathrm{K}$, $k_0 = 0.2~{\rm eV/(\AA^2 u)}$ and 
$k=1~{\rm eV/(\AA^2 u)}$.
\label{fig2}}
\end{figure}

Next, we consider the effect of Coriolis force on the steady-state currents $I_{\alpha}$ Eq. (\ref{eq:Landauer-like}) flowing out of 
each of the three baths (a negative current indicates heat flowing towards the bath). From the calculations we confirm that a current 
conservation is satisfied as expected i.e. $I_{R_1}=-I_{R_{2}}-I_{R_{3}}$ . A typical dependence of the heat currents on the angular 
velocity $\Omega$, for a fixed angle $\theta=60^\circ$, is shown in Fig.~\ref{fig2}. We have assumed CCW rotation of the platform 
while $T_{R_1}=T_H$ and $T_{R_2}=T_{R_3}=T_L$ (see Fig. \ref{fig1}). Despite the geometric symmetry of the set-up, we find that 
$-I_{R_3}\gg -I_{R_2}$. We have also confirmed via direct calculations of the currents $I_{\alpha}$ (not shown here) that this behaviour 
is insensitive to the presence of the direct coupling between particles $R_2$ and $R_3$. Based on symmetry considerations we 
further conclude that a re-arrangement of the bath temperatures such that $T_{R_3}=T_H$ and $T_{R_2}=T_{R_1}=T_L$, will lead 
to a rectification effect which favors the heat current flowing towards bath $R_2$ i.e. $-I_{R_2}\gg -I_{R_1}$. Likewise, the temperature 
configuration $T_{R_2}=T_H, T_{R_1}=T_{R_3}=T_L$, leads to a rectified current towards the particle $R_1$ i.e. $-I_{R_1}\gg -I_{R_3}$. 
The CCW current propagation is reminiscent of the operation of a circulator - a device used in microwaves. 

The origin of the asymmetric heat current flow can be further traced to the non-reciprocal transmission between various baths i.e
${\cal T}_{\gamma,\alpha}(\omega,\Omega)\neq {\cal T}_{\alpha,\gamma}(\omega,\Omega)$ which enter the expression Eq.~ 
(\ref{eq:Landauer-like}) for the heat currents $I_{\alpha}$. In fact, one can prove that the following relation holds ${\cal T}_{\gamma,
\alpha}(\omega,\Omega)= {\cal T}_{\alpha,\gamma}(\omega,-\Omega)$ \cite{KLK15}.

\begin{figure}
\includegraphics[width=1\linewidth,angle=0]{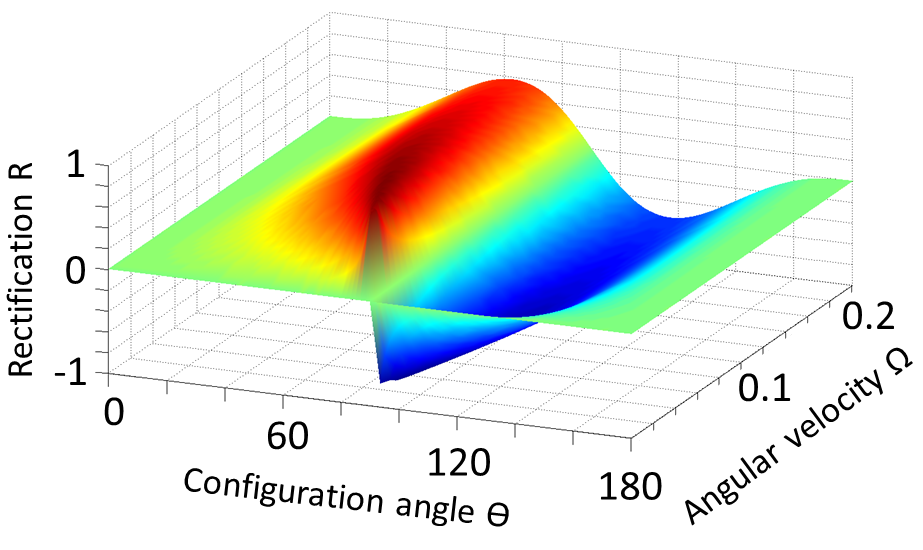}
\caption{(Color online) 3D plot of rectification parameter $R$ versus  configuration angle $\theta$ [degrees] and angular velocity 
$\Omega \left[{\rm \frac{1}{\AA}\sqrt{\frac{eV}{u}}}\right]$.Other parameters are $T_{R_1}=T_{H}=310\mathrm{K},\: T_{R_2}=
T_{R_3}=T_{L}=290\mathrm{K}$, $k_0 = 0.2~{\rm eV/(\AA^2 u)}$ and $k=1~{\rm eV/(\AA^2 u)}$.
\label{fig3}}
\end{figure}

To quantify the thermal rectification effect, we introduce a rectification parameter ${\cal R}$, which is defined 
as 
\begin{equation}
\label{rectif}
{\cal R}\equiv\frac{I_{R_{2}}-I_{R_{3}}}{-I_{R_{2}}-I_{R_{3}}}=\frac{I_{R_{2}}-I_{R_{3}}}{I_{R_1}}.
\end{equation}
where we have assumed that $T_{R_1}=T_H$ and $T_{R_2}=T_{R_3}=T_L$ (similar definitions can be used in case of different arrangement 
of the baths). Generally, ${\cal R}$ is a function of the angular velocity $\Omega$ and the equilibrium configuration 
angle $\theta$ (see Fig.~\ref{fig1}) and it takes values between ${\cal R}\in [-1,1]$. The two extreme limits $\pm 1$ corresponds 
to maximal asymmetry in the heat current towards reservoirs $R_3$ or  $R_2$ respectively. The case ${\cal R}=0$ correspond to symmetric 
heat flow towards the two cold reservoirs. 

A panorama of the dependence of ${\cal R}(\Omega,\theta)$ on $\theta$ and $\Omega$ is shown in Fig. \ref{fig3}. A feature of this 
analysis is that above a critical angle $\theta^*=90^\circ$ the rectification parameter change sign indicating that a CCW rotation of 
the platform will result to a current rectification effect towards the cold bath $R_2$ which is next to the hot one $R_1$ in the clockwise 
direction. Moreover we see that there is an optimal configuration angle $\theta\approx 85^\circ$ ($\theta\approx 95^\circ$) for 
which the rectification parameter gets its extreme values ${\cal R}\approx 0.9 ({\cal R}\approx -0.9)$ for a critical value of rotation
angle $\Omega\approx 0.02\ll \omega_n$. Because the later is externally controlled, our set-up provides a high degree of tunability, 
with the possibility of changing from reciprocal ($\Omega\approx 0$) to nonreciprocal ($\Omega\neq 0$) behaviour. Moreover we 
can reverse the handedness of the circulator, say from CCW to CW, by simply changing the rotation direction of the platform.

{\it Conclusions -} We have introduced the concept of Coriolis linear circulators which employ the Coriolis force in order to break 
time-reversal symmetry in the circulating flow of heat currents. We have validated our proposal via theoretical calculations with a 
simple model consisting of three mutually coupled harmonic masses which move on a rotating frame with angular velocity $\Omega$. 
The efficiency of the rectification effect depends on the magnitude of the angular velocity. Surprisingly optimal rectification can be 
achieved for moderate values of $\Omega$. It will be interesting to promote and investigate the efficiency of this proposal to realistic 
set-ups. An example case is a rotating graphene flake where the contact with the baths can be achieved optically, via optical heating 
and cooling. Some interesting questions along this line include the persistance of Coriolis thermal rectification beyond the ballistic 
(e.g. to diffusive) transport regime, the effects of lattice defects in rectification, at the form of the thermal current counting statistics 
in the presence of Coriolis forces \cite{KLK15}. 


{\it Acknowledgments-}We acknowledge useful discussions and suggestions from F. Ellis and T. Prosen who participated at the initial 
phase of this project. This work was partly sponsored by a NSF DMR-1306984 grant and by an AFOSR MURI grant FA9550-14-1-0037.

\end{document}